\documentclass[doublecol,linenumbers]{epl2}
\usepackage{amssymb,latexsym,amsmath,amscd}
\usepackage{color,soul}
 \title{Polarizable polymer chain under external electric field: Effects of many-body electrostatic dipole correlations}
  \shorttitle{Polarizable polymer chain under electric field}
\author{Yu.~A. Budkov\inst{1} \thanks{E-mail: \email{ybudkov@hse.ru}} \and A. L. Kolesnikov\inst{2} \thanks{E-mail: \email{bancocker@mail.ru}}}
\shortauthor{Yu. A. Budkov \etal}

\institute{
  \inst{1} Department of Applied Mathematics, National Research University Higher School of Economics, Moscow, Russia\\
	\inst{2} Institut f\"{u}r Nichtklassische Chemie e.V., Universit\"{a}t Leipzig - Leipzig, Germany\\}


\abstract{We present a new simple self-consistent field theory of the polarizable flexible polymer chain under the external constant electric field with account for the many-body electrostatic dipole correlations. We show the effect of electrostatic dipole correlations on the electric field-induced globule-coil transition. We demonstrate that only when the polymer chain is in the coil conformation, the electrostatic dipole correlations of monomers can be considered as pairwise. However, when the polymer chain is in collapsed state, the dipole correlations have to be considered at the many-body level.}

\begin{document}
\maketitle

\section{Introduction}
Nowadays statistical thermodynamics of dielectric polymers is one of the most unexplored areas of polymer physics. Indeed, till now only several
theoretical works have been published, where thermodynamic and structural properties of dielectric polymers in the bulk solution
\cite{Podgornik_2004,Kumar_2009,Dean_2012,Kumar_2014,Lu2015} and under external electric field \cite{Budkov_2015} have been discussed.
In ref. \cite{Podgornik_2004} Podgornik investigated within the path integrals formalism the renormalized persistence length of semi-flexible polymer
chain whose segments interact via a screened Debye-H\"{u}ckel dipolar interaction potential. In ref. \cite{Kumar_2009} Kumar et al. within the
Edwards-Singh method calculated the mean-square radius of gyration of polyzwitterionic molecules in aqueous solutions as a function of the chain
length, electrostatic interaction strength, added salt concentration, dipole moment, and degree of ionization of the zwitterionic monomers. In ref.
\cite{Dean_2012} Dean et al. showed that taking into account the polarizing  many-body correlations at the level of random phase approximation (RPA)
can lead to ordering of the semi-flexible anisotropic polymer chains in the solution. In ref. \cite{Kumar_2014} Kumar et al. showed within the
field-theoretic formalism that dipolar interactions in polymer blends can significantly enhance the phase segregation. Lu et al. within the
field-theoretic formalism analyzed the van der Waals interactions between two rigid polymers polarizable along their backbone \cite{Lu2015}.

In our recent work \cite{Budkov_2015} we investigated the conformational behavior of the polarizable flexible polymer chain under the external
electric field within the Flory-type mean-field theory \cite{Flory_book}. We showed that regardless the polymer chain conformation (coil or globule)
electric field increase causes the swelling of polymer chain. We also showed that increasing the electric field in the regime of poor solvent can
provoke the globule-coil transition. However, we fully neglected the many-body electrostatic correlations between monomers related to their molecular
polarizability which cannot be accounted at the level of pure mean-field theory \cite{Dean_2012}. These polarizing correlations, related to the fluctuations of local electrostatic potential, could be accounted via the fluctuation corrections to the mean-field approximation. It is evident in
advance that contribution of the latter effects to the total free energy must be important at enough large monomer polarizability. Thus, the
natural question appears: {\sl How the electrostatic many-body correlations of monomers can change the polarizable polymer chain
conformational behavior under the external electric field?}

\section{Theory}
Let us consider a polarizable flexible polymer chain immersed in a dielectric solvent which we model as a continuous dielectric medium with the
dielectric permittivity $\varepsilon_{s}$.  Let polymer chain has a degree of polymerization $N$ and each monomer has a molecular polarizability
$\gamma$. The monomer polarizability may be related to the electronic polarizability of monomers as well as the orientational polarizability of their permanent dipoles. The first case can be realized for the synthetic glassy polymers. The second case of permanent monomer dipoles is possible for the weak polyelectrolytes in the regime of counterion condensation, when the counterions and monomers form the strongly bound ion pairs. We consider only the case of isotropic dielectric response for simplicity. We also assume that the polymer chain is under the homogeneous electric field $\bold{E}$. To study the conformations of the polymer chain in the external electric field, we use the simple Flory-type \cite{Flory_book} mean-field theory, considering the radius of gyration $R_{g}$ as a single order parameter. Therefore, we assume that the polymer chain occupies the volume which can be estimated by the gyration volume $V_{g}=4\pi R_{g}^3/3$.  Using all above-mentioned model assumptions, one can write a total free energy of the polymer chain in the following form:
\begin{equation}
\mathcal{F}(R_{g})=\mathcal{F}_{conf}(R_{g})+\mathcal{F}_{vol}(R_{g})+\mathcal{F}_{el}(R_{g}),
\end{equation}
where $\mathcal{F}_{conf}(R_{g})$ is the free energy of the ideal polymer chain which can be calculated by the following interpolation formula
\cite{Fixman,Grosberg,Budkov1,Budkov,Budkov2,Budkov3,Budkov4}
\begin{equation}
\label{eq:conf}
\mathcal{F}_{conf}(R_{g})=\frac{9}{4}k_{B}T\left(\alpha^{2}+\frac{1}{\alpha^2}\right),
\end{equation}
where $\alpha=R_{g}/R_{0g}$  is the expansion factor, $R_{0g}^2=Nb^2/6$ is the mean-square radius of gyration of the Gaussian polymer chain, $b$ is
the Kuhn length, $k_{B}$ is the Boltzmann constant, $T$ is the temperature; the contribution of pairwise interactions of monomers to the total free energy can be accounted for simplicity via the standard virial series as follows
\begin{equation}
\label{eq:vol}
\mathcal{F}_{vol}(R_{g})=k_{B}T\left(\frac{N^2B}{2V_{g}}+\frac{N^3C}{6V_{g}^2}\right),
\end{equation}
where $B$ and $C$ are the second and third virial coefficients, respectively; the electrostatic contribution can be written as a sum of two
contributions
\begin{equation}
\label{eq:electr}
\mathcal{F}_{el}(R_{g})=\mathcal{F}_{el}^{(MF)}(R_{g})+\mathcal{F}_{el}^{(fl)}(R_{g}),
\end{equation}
where first term is a mean-field approximation for the electrostatic free energy which can be estimated as the free energy of dielectric sphere
\cite{Landau_VIII}
\begin{equation}
\label{eq:electr1}
\mathcal{F}_{el}^{(MF)}(R_{g})=-\frac{V_{g}E^2}{8\pi}\frac{3\varepsilon_{s}\left(\varepsilon_{p}-\varepsilon_{s}\right)}{2\varepsilon_{s}+\varepsilon_{p}}=-\frac{3N\varepsilon_{s}\gamma
E^2}{2 \left(3\varepsilon_{s}+\frac{4\pi \gamma N}{V_{g}}\right)},
\end{equation}
where the effective dielectric permittivity inside the polymer volume $\varepsilon_{p}=\varepsilon_{s}+{4\pi\gamma N}/{V_{g}}$ in the mean-field approximation is introduced \cite{Budkov_2015,Budkov2015_2}; $\gamma$ is the molecular polarizability of monomers, and $E=|\bold{E}|$ is the absolute value of the external electric field. The second term in eq. (\ref{eq:electr}) determines the contribution of correlations between fluctuating dipoles. This contribution can be assessed for the enough large polymer volume within the formalism proposed in \cite{Dean_2012} at the level of random phase approximation (RPA) for the case of isotropic dielectric response:
\begin{equation}
\label{eq:electr2}
\mathcal{F}_{el}^{(fl)}\simeq\frac{V_{g}k_{B}T}{2}\int\limits_{|\bold{k}|<\Lambda}\frac{d\bold{k}}{(2\pi)^3}\ln\left(\frac{V_{s}(\bold{k})}{V_{p}(\bold{k})}\right)\nonumber
\end{equation}
\begin{equation}
=\frac{2\pi
V_{g}k_{B}T}{3b^3}\ln\left(1+\frac{4\pi\gamma }{\varepsilon_{s}}\frac{N}{V_{g}}\right),
\end{equation}
where $V_{p}(\bold{k})=4\pi/(\varepsilon_{p}\bold{k}^2)$ and $V_{s}(\bold{k})=4\pi/(\varepsilon_{s}\bold{k}^2)$ are the Fourier-images of Coulomb potentials inside the polymer volume and in the pure solvent, respectively; $\Lambda=2\pi/b$ is the parameter of ultraviolet cut-off. The choice of such value of the cut-off parameter $\Lambda$ is due to the fact that at the scales $\sim b$ fluctuations of the electrostatic potential related to the dipoles fluctuations are absent \cite{Dean_2012}.

Collecting together all above mentioned expressions, we arrive at the following result for the total free energy of the polymer chain
in the solution under external electric field:
\begin{equation}
\label{eq:total}
\frac{\mathcal{F}}{k_{B}T}=\frac{9}{4}\left(\alpha^{2}+\frac{1}{\alpha^2}\right)+\frac{N^2B}{2V_{g}}+\frac{N^3C}{6V_{g}^2}\nonumber
\end{equation}
\begin{equation}
-\frac{3N\varepsilon_{s}\gamma
E^2}{2k_{B}T\left(3\varepsilon_{s}+\frac{4\pi\gamma N}{V_{g}}\right)}+\frac{2\pi V_{g}}{3b^3}\ln\left(1+\frac{4\pi\tilde{\gamma}}{\varepsilon_{s}}\frac{N}{V_{g}}\right).
\end{equation}
We would like to stress that in our previous work \cite{Budkov_2015} we have estimated the mean-field electrostatic contribution as the free energy of
dielectric plate, whereas here we use more appropriate relation for the free energy of dielectric sphere.

\section{Numerical results and discussions}
To perform some analytical estimates and numerical calculations in further, it is convenient to introduce the following dimensionless variables:
$\tilde{E}=E\sqrt{\varepsilon_{s}b^3/k_{B}T}$, $\tilde{B}=Bb^{-3}$, $\tilde{C}=Cb^{-6}$, and $\tilde{\gamma}=\gamma b^{-3}/\varepsilon_{s}$.

Further, minimizing the total free energy (\ref{eq:total}) with respect to the expansion factor $\alpha$, after some algebra we arrive at the equation in dimensionless form
\begin{equation}
\label{eq:alpha}
\alpha^5-\alpha=\frac{3\sqrt{6}}{2\pi
}\tilde{B}\sqrt{N}+\frac{27\tilde{C}}{\pi^2\alpha^3}+\frac{2\sqrt{6}\tilde{\gamma}^2\sqrt{N}\tilde{E}^2}{\left(1+\frac{6\sqrt{6}\tilde{\gamma}}
{\alpha^3\sqrt{N}}\right)^2}-\nonumber
\end{equation}
\begin{equation}
\frac{4\sqrt{6}\pi^2N^{3/2}\alpha^6}{243}\left(\ln\left(1+\frac{18\sqrt{6}\tilde{\gamma}}{\alpha^3\sqrt{N}}\right)-
\frac{\frac{18\sqrt{6}\tilde{\gamma} }{\alpha^3\sqrt{N}}}{1+\frac{18\sqrt{6}\tilde{\gamma}}{\alpha^3\sqrt{N}}}\right).
\end{equation}
The first and second terms in the right-hand side of eq. (\ref{eq:alpha}) determine the effect of volume interactions. The third term determines the
influence of the interactions of induced dipoles with the applied electric field on the polymer swelling. As is seen, the latter always leads to
swelling of the polymer chain that is related to the well known electrostriction phenomena \cite{Landau_VIII}. The forth term determines the effect of
many-body correlations of the fluctuating dipoles on the polymer swelling.

To understand how the electrostatic dipole correlations can affect the polymer conformation, at first we consider the case of repulsive volume
interactions ($\tilde{B}>0$), when the polymer chain is in conformation of expanded coil ($\alpha\gg 1$). Thus in this limit we get
\begin{equation}
\label{eq:alpha2}
\alpha^5-\alpha\simeq\frac{3\sqrt{6}}{2\pi}\sqrt{N}\left(\tilde{B}+\frac{4\pi}{3}\tilde{\gamma}^2\tilde{E}^2-\frac{32\pi^3}{3}\tilde{\gamma}^2\right).
\end{equation}
The equation (\ref{eq:alpha2}) means that correlations between fluctuating dipoles in the coil state lead to effective decreasing in the second virial
coefficient. In the case when the monomer polarizability is attributed to the orientational polarizability of permanent dipoles, i.e., when $\gamma=p^2/(3k_{B}T)$ ($p$ is a permanent dipole), we get the following equation for the expansion factor at $\tilde{E}=0$
\begin{equation}
\label{eq:alpha3}
\alpha^5-\alpha\simeq\frac{3\sqrt{6}}{2\pi b^3}\sqrt{N}\left(B-\frac{32\pi^3p^4}{27(k_{B}T)^2\varepsilon_{s}^2b^3}\right).
\end{equation}
It should be noted that eq.(\ref{eq:alpha3}) is similar to that was obtained for the weak polyelectrolyte chain in the regime of counterion condensation \cite{Dua2014}. The latter equation means that when the polymer chain is in coil conformation, the electrostatic dipole correlations can be considered as pairwise. However, when the polymer chain adopts a collapsed state, the higher dipole correlations become important. In other words, when the polymer chain is in the globule state, polarizing dipole correlations have to be accounted at the many-body level.

In order to elucidate a role of the many-body dipole correlations in the polymer chain conformational behavior, let us consider the dependence of the expansion factor on the monomer polarizability $\tilde{\gamma}$ at zero electric field. We assume in this case that the volume interactions are repulsive, i.e., that $\tilde{B}>0$. On fig. 1 are depicted the dependences of the expansion factor on the monomer polarizability at the different second virial coefficients $\tilde{B}$. As is seen, increasing in the monomer polarizability leads to the dramatic decrease in the expansion factor. The latter result means that the enough strong electrostatic many-body correlations of monomers can provoke the coil-globule transition. It should be noted that this effect is quite similar to the coil-globule transition of polyelectrolyte chain induced by correlations of charges \cite{DeLaCruz2000,Brilliantov_1998,Brilliantov_2016,Pincus,Cherstvy2010,Netz_2003,Netz_2003_2}. It is worth noting that in the theory (see, for instance, ref. \cite{Dua2014}), where the monomer dipole correlations were considered as pairwise, in order to compensate the attraction of the dipoles, it is needed to take into account the repulsive volume interactions up to the third term of the virial expansion. However, accounting for the dipole correlations at the many-body level, as is seen from fig. 1, allows compensate the attraction of the dipoles taking into account the repulsive volume interactions even at the level of second virial coefficient.

\begin{figure}
 \centerline{\includegraphics[scale = 0.95]{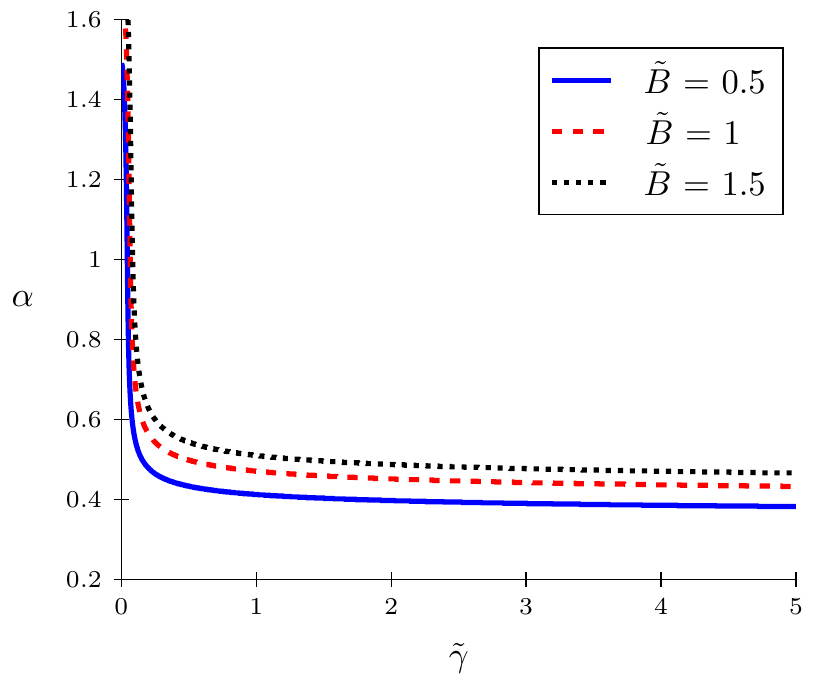}}
 \caption{The expansion factor $\alpha$ as a function of the monomer polarizability at the different second virial coefficients $\tilde{B}$.
 The data are shown for $N=100$, $\tilde{E}=0$, $\tilde{C}=0$.}
 \label{fig.1}
 \end{figure}

\begin{figure}
\centerline{\includegraphics[scale = 0.95]{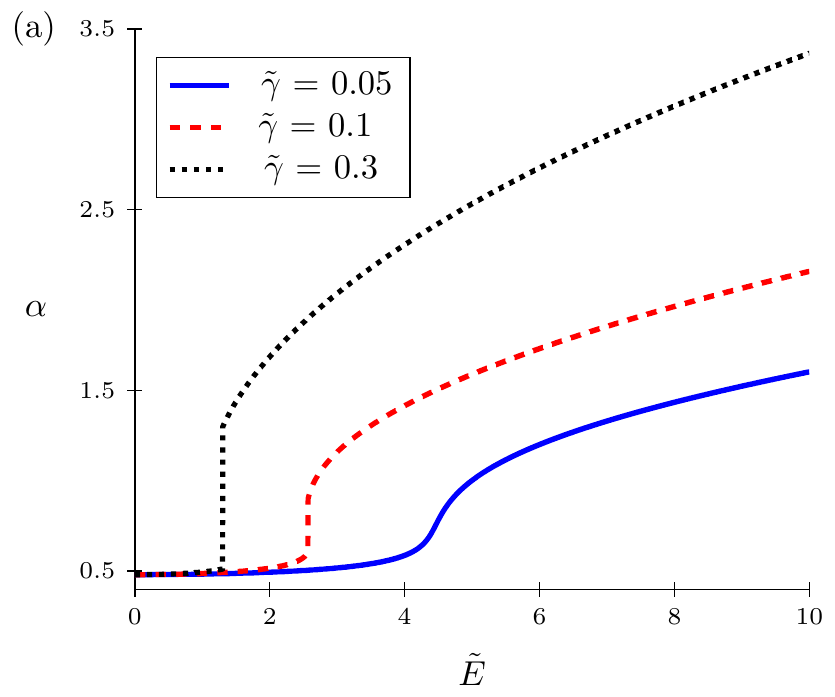}}
\centerline{\includegraphics[scale = 0.95]{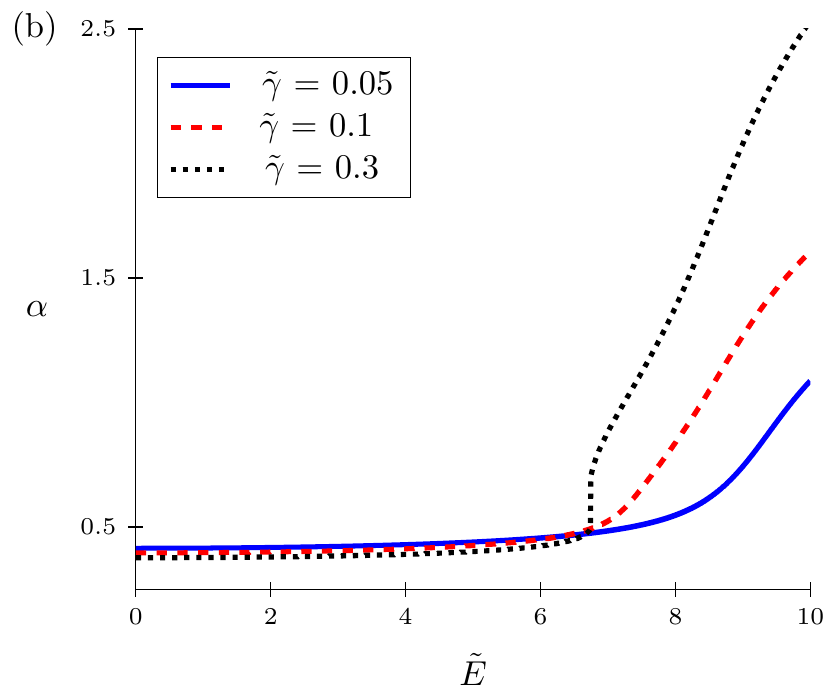}}
\caption{The expansion factor $\alpha$ as a function of the electric field $\tilde{E}$ is calculated by the (a) mean-field theory and (b) theory
with account for the electrostatic dipole correlations at the different monomer polarizabilities $\tilde{\gamma}$.
The data are shown for $N=100$, $\tilde{B}=-0.25$, $\tilde{C}=0.1$.}
\label{fig.1}
\end{figure}

As one can see from eq. (\ref{eq:alpha}), the presence of electric field inside the polymer coil,
oppositely, leads to an effective increase in the second virial coefficient. In the case of strong electric field ($\tilde{E}\gg 1$), we get the
limiting laws for the expansion factor and the radius of gyration:
\begin{equation}
\alpha\sim \tilde{\gamma}^{2/5}\tilde{E}^{2/5}N^{1/10}, ~~ R_{g}/b\sim \tilde{\gamma}^{2/5}\tilde{E}^{2/5}N^{3/5},
\end{equation}
which were first obtained (up to numerical prefactors) and discussed in ref. \cite{Budkov_2015}.

Now let us pass to the discussion of how the many-body electrostatic correlations of the fluctuating dipoles can change the conformation behavior of
the polymer chain under the external electric field compared to the pure mean-field theory. Especially, we would like to focus on the influence
of the electrostatic correlations of monomers on the {\sl electric field-induced globule-coil transition} which was in detail discussed in ref.
\cite{Budkov_2015} in the framework of mean-field theory. We assume that $\tilde{B}=-0.25$ and $\tilde{C}=0.1$, so that polymer chain is in the
globule state even at $\tilde{\gamma}=0$. Fig. 2 demonstrates the expansion factor as a function of the electric field at the different monomer
polarizability $\tilde{\gamma}$ obtained within the (a) pure mean-field theory and (b) present theory with accounting for the many-body electrostatic dipole correlations. As is seen, in both cases applying the electric field exceeding some threshold value induces the
globule-coil transition. Nevertheless, account for the electrostatic correlations shifts this transition to larger electric fields. It is worth
noting that in the region of small electric fields, when the polymer chain is in collapsed state, electrostatic correlations lead to
smaller values of the expansion factor than that are predicted by the mean-field theory. Thereby this phenomenon is reminiscent
the globule-coil transition of the polyelectrolyte chain caused by the electic field \cite{Netz_2003,Netz_2003_2}. It should be noted that in the
region of sufficiently large monomer polarizability in both theories globule-coil transition occurs as a first-order phase
transition, i.e., as abrupt increase in the expansion factor.

\section{Conclusion}
In conclusion, we have formulated the simple Flory-type self-consistent field theory of the polarizable polymer chain under the external
electric field with account for the many-body dipole electrostatic correlations.  We have shown that when the polymer chain is in the coil state,
while the monomer polarizability is small, the electrostatic dipole correlations can be considered as pairwise. In this case their
effect consists of the decrease in the second virial coefficient of monomer-monomer interaction. However, at enough strong monomer
polarizability, electrostatic dipole correlations can cause the coil-globule transition. When the polymer chain is in the globule state,
electrostatic dipole correlations have to be considered at the many-body level. We have also shown that the account for the many-body electrostatic dipole
correlations does not qualitatively change the main result of our previous pure mean-field theory \cite{Budkov_2015} --
the {\sl electric field-induced globule-coil transition}. However, in the present theory the electric field, at which
the globule-coil transition takes place, shifts to higher values. Such trend is related to the fact that an availability of the polarizability on the
polymer backbone leads to the additional effective attraction between monomers. So it requires to apply stronger electric field to disjoin the polymer
globule.
\begin{acknowledgments}
This work was supported by grant from Russian Foundation for Basic Research (Grant No. 15-43-03195).
\end{acknowledgments}


\begin{thebibliography}{99}
\bibitem{Podgornik_2004}
{Rudi Podgornik}, Phys. Rev. E, $\bold{70}$ (2004) 031801.
\bibitem{Kumar_2009}
{Rajeev Kumar and Glenn H. Fredrickson}, J. Chem. Phys., $\bold{131}$ (2009) 104901.

\bibitem{Dean_2012}
{David S. Dean and Rudolf Podgornik},  J. Chem. Phys., $\bold{136}$ (2012) 154905.

\bibitem{Kumar_2014}
{Rajeev Kumar, Bobby G. Sumpter, and M. Muthukumar}, Macromolecules, $\bold{47}$ (2014) 6491.

\bibitem{Lu2015}
{Bing-Sui Lu, Ali Naji, and Rudolf Podgornik}, J. Chem. Phys., $\bold{142}$ (2015) 214904.

\bibitem{Budkov_2015}
{Budkov Yu.A., Kolesnikov A.L., Kiselev M.G.}, J. Chem. Phys., $\bold{143}$ (2015) 201102.

\bibitem{Flory_book}
{Flory P.} {\sl Statistical Mechanics of Chain Molecules} (New York, Wiley-Interscience) 1969.

\bibitem{Fixman}
{Fixman M.}, J. Chem. Phys., $\bold{36}$ (2) (1962) 306.

\bibitem{Grosberg}
{Grosberg A.Yu., Kuznetsov D.V.},  Macromolecules, $\bold{25}$ (1992) 1970.

\bibitem{Budkov1}
{Budkov Yu.A., Kolesnikov A.L., Georgi N., and Kiselev M.G.} J. Chem. Phys. $\bold{141}$, 014902 (2014).

\bibitem{Budkov}
{Yu. A. Budkov, I. I. Vyalov, A. L. Kolesnikov, N. Georgi, G. N. Chuev, and M. G. Kiselev} J. Chem. Phys. $\bold{141}$, 204904 (2014).

\bibitem{Budkov2}
{Budkov Yu.A., Kolesnikov A.L., Georgi N., Kiselev M.G.} Euro. Phys. Lett. $\bold{109}$, 36005 (2015).

\bibitem{Budkov3}
{Budkov Yu.A., Kolesnikov A.L., Kiselev M.G.} J. Chem. Phys. $\bold{143}$, 201102 (2015).

\bibitem{Budkov4}
{Budkov Yu. A., Kolesnikov A. L., Kalikin N. N., and Kiselev M. G.} Euro. Phys. Lett., $\bold{114}$, 46004 (2016).

\bibitem{Landau_VIII}
{Landau L.D., Lifshitz E.M.}  {\sl Electrodynamics of Continuous Media V. 8, A Course of Theoretical Physics} (Pergamon Press, Oxford, UK) 1960.

\bibitem{Budkov2015_2}
{Yu. A. Budkov, A. L. Kolesnikov, and M. G. Kiselev}, Europhys. Lett., 111 (2015) 28002.

\bibitem{DeLaCruz2000}
{Francisco J. Solisa and Monica Olvera de la Cruz} J. Chem. Phys., $\bold{112}$ (2000) 2030.

\bibitem{Brilliantov_1998}
{Brilliantov N.V., Kuznetsov D.V., Klein R.}, Phys. Rev. Lett., $\bold{81}$ (7) (1998) 1433.

\bibitem{Brilliantov_2016}
{Anvy Moly Tom, Satyavani Vemparala, R. Rajesh, and Nikolai V. Brilliantov}, Phys. Rev. Lett., $\bold{117}$ (2016) 147801.

\bibitem{Pincus}
{Schiessel H., Pincus P.}, Macromolecules, $\bold{31}$ (1998) 7953.

\bibitem{Cherstvy2010}
{A. G. Cherstvy}, J. Phys. Chem. B, $\bold{114}$ (16) (2010) 5241.

\bibitem{Netz_2003}
{Netz R.R.}, J. Phys. Chem. B, $\bold{107}$ (2003) 8208.

\bibitem{Netz_2003_2}
{R. R. Netz}, Phys. Rev. Lett., $\bold{90}$ (2003) 128104.

\bibitem{Dua2014}
{Prasanta Kundu and Arti Dua} Journal of Statistical Mechanics: Theory and Experiment, $\bold{2014}$ (2014) 07023.

\end{thebibliography}
 \end{document}